Improvement of the solubilization of proteins in two-dimensional electrophoresis with immobilized pH gradients.

Thierry Rabilloud [1*], Céline Adessi [2], Anne Giraudel [3] and Joël Lunardi [1]

1: CEA- Laboratoire de Bioénergétique Cellulaire et Pathologique, UA 2019
DBMS/BECP
CEA-Grenoble, 17 rue des martyrs
F-38054 GRENOBLE CEDEX 9 FRANCE

2: CEA - Laboratoire de Chimie des Protéines,
DBMS / CP
CEA-Grenoble, 17 rue des martyrs
F-38054 GRENOBLE CEDEX 9 FRANCE

3: CEA - Laboratoire du Cytosquelette, Unité INSERM U 366
DBMS / CS
CEA-Grenoble, 17 rue des martyrs
F-38054 GRENOBLE CEDEX 9 FRANCE

(Running title): two-dimensional electrophoresis of sparingly soluble proteins

*: to whom correspondence should be addressed

Correspondence :
Thierry Rabilloud, DBMS/BECP
CEA-Grenoble, 17 rue des martyrs,
F-38054 GRENOBLE CEDEX 9
Tel (33)-76-88-32-12
Fax (33)-76-88-51-87
e-mail: Thierry@sanrafael.ceng.cea.fr

Summary

We have carried out the separation of sparingly-soluble (membrane and nuclear) proteins by high resolution two-dimensional electrophoresis. IEF with immobilized pH gradients leads to severe quantitative losses of proteins in the resulting 2-D map, although the resolution is usually kept high. We therefore tried to improve the solubility of proteins in this technique, by using denaturing cocktails containing various detergents and chaotropes. Best results were obtained by using a denaturing solution containing urea, thiourea, and detergents (both nonionic and zwitterionic). The usefulness of thiourea-containing denaturing mixtures are shown in this article on several models including microsomal and nuclear proteins and on tubulin, a protein highly prone to aggregation.

Abbreviations

APS: ammonium persulfate; BCA: bicinchoninic acid; Bis: N-N' methylene bis acrylamide; BisTris: N-N bis (hydroxyethyl) N-N-N tris (hydroxymethyl) aminomethane; %C: ratio (in percent) of crosslinker to total monomers; CA-IEF: isoelectric focusing in pH gradients generated by carrier ampholytes; CHAPS: 3-[(3-cholamidopropyl) dimethylammonio] propane sulfonate; DMEM: Dulbecco Modified Eagle's medium; DTT: dithiothreitol; EtOH: ethanol; HEPES: 4-(2-hydroxyethyl) 1 piperazine ethane sulfonic acid; IAM: iodoacetamide; IEF: Isoelectric focusing; IPG: IEF in immobilized pH gradients; IPG-DALT: 2-D electrophoresis with IEF in immobilized pH gradient in the first dimension and SDS PAGE in the second dimension; PAGE: polyacrylamide gel electrophoresis; PBS: phosphate buffered saline; PDA: 1,4-bis(acryloyl) piperazine; Pefabloc: aminoethyl benzene sulfonyl fluoride; PMSF: phenyl methyl sulfonyl fluoride; RUBISCO: ribulose 1,5 biphosphate carboxykinase; SDS: sodium dodecyl sulfate; %T: grams of monomers (including crosslinker) per 100 ml of gel; TCA: trichloroacetic acid; Temed: tetramethyl ethylene diamine; Tris: Tris(hydroxymethyl)aminomethane

INTRODUCTION

Since its indroduction in the mid seventies, high resolution 2-D electrophoresis has been used for many types of proteins, including membrane proteins (e.g. [1-5]). However, conventional isoelectric focusing, where the pH gradient is established with carrier ampholytes is plagued by pH gradient instability with time and irreproducibility on the long term. Both problems are mainly due to the undefined chemical composition of carrier ampholytes. The use of immobilized pH gradients for the focusing step [6, 7] has eliminated these drawbacks. In this system, the pH gradient is built by a small number of defined chemicals and covalently grafted to the polyacrylamide matrix. This allows true steady-state focusing, which greatly increases reproducibility. Additional advantages of immobilized pH gradients include a high capacity [8], which allows preparative runs in which microgram amounts of proteins are available for structural analysis [9, 10].

Separation of membrane proteins has been shown to be feasible [11, 12], and 2-D electrophoresis of membrane proteins with immobilized pH gradients has been presented on some occasions [13], with adequate resolution. However, it was recently shown that severe quantitative losses were experienced when membrane proteins were separated by IPG-DALT [14]. The severity of these losses increased with the protein load, so that micropreparative gels of membrane fractions yielded correct amounts of proteins mainly for the soluble proteins contaminating the preparation (C. Miege, M. Vinçon, T. Rabilloud, unpublished results). These losses were attributed to adsorption of proteins on the immobilized pH gradient matrix at (or close to) the isoelectric point [14].

In order to alleviate this problem, we investigated the denaturing cocktail used for IEF with immobilized pH gradients, and we found that a proper combination of chaotropes and detergents greatly improved the solubility of the proteins in the first dimension and their transfer into the SDS gel. The key factors seemed to be the presence of thiourea in addition to urea as a chaotrope and the presence of zwitterionic amphiphilic compounds (e.g. CHAPS or SB 3-10) in the denaturing solution.

MATERIAL AND METHODS

Chemicals and materials

Immobilines, Pharmalytes 3-10, gel cassettes, gradient maker and the equipment for running the IPG gels (Multiphor II and Dry-Strip kit) were from Pharmacia. Resolytes 4-8 were from BDH. Second-dimension SDS gels were cast and run in a Bio-Rad Multi Cell. PDA and protein assay reagent were from Bio-Rad, BCA protein assay reagents, Pefabloc and Hepes hemisodium salt from Sigma, and acrylamide from Amresco. All the other chemicals were analytical grade from Fluka. Gel-Bond PAG film was from FMC and Gel Fix Covers from Serva.

Protein sample preparation
Chloroplast envelopes

Chloroplast envelopes from spinach (a gift from C. Miege and C. Albrieux) were prepared according to Douce and Joyard [15]. The envelopes were pelleted at 100000g for 1 hour, the resulting pellet was dissolved in the required denaturing solution, and the resulting solution was directly used for 2-D electrophoresis experiments.

Microsomal and nuclear proteins

C2C12 mouse embryonic myoblasts were cultured in DMEM (4.5g glucose per liter) + 10% fetal calf serum on gelatinized culture vessels, while P3-X63-Ag8 plasmocytoma cells were cultured in suspension in RPMI1640 medium containing 1mM pyruvate, 10μM mercaptoethanol and 10% fetal calf serum. Cells were harvested by scraping and resuspended in isotonic buffer A (Hepes-NaOH 10mM pH 7.5, sucrose 0.25M, EGTA 1mM, fatty acid-free BSA 0.5%). The cells were broken with a Dounce homogeneizer. A first centrifugation at 1000g for 5 minutes led to a pellet of crude nuclei. The supernatant was centrifuged at 10000g for 10 minutes to pellet the mitochondria and then at 100000g for 1 hour to pellet the microsomes. The microsomal fraction was resuspended in buffer A and the protein concentration determined by a microBCA assay in the presence of SDS [16]. For use in 2-D electrophoresis the required amount of microsome suspension was diluted in 1 ml of PBS, centrifuged at 100000g for 1 hour. The resulting microsome pellet was dissolved in the required denaturing solution and used directly in 2-D electrophoresis experiments.

The pellet of crude nuclei was cleaned by resuspension in buffer B (Hepes-NaOH 10mM pH 7.5, sucrose 0.34M, spermidine 0.75 mM, spermine 0.15mM, EDTA 1mM, Triton X100 0.1%) and recentrifugation at 1000g for 5 minutes. The purified nuclei were lysed in 10 volumes of solution C (urea 9M, CHAPS 4%, spermine base 10mM, DTT 40mM) for 1 hour at room temperature. The highly viscous lysate was centrifuged at 200000g for 1 hour to pellet the nucleic acids. The supernatant was recovered and its protein concentration determined by a Bradford assay [17], using BSA as a standard. Carrier ampholytes (Pharmalytes 3-10) were added at a final concentration of 0.4%, and the solution was used immediately or kept frozen at -20°C. For use in 2-D electrophoresis, this concentrated solution was diluted in the required denaturing solution.

Integral membrane proteins

*D. discoideum* strain AX2 (ATCC 24397) amoebae ($1.10^8$ cells) were washed three times in a ice-cold 40 mM Mes-Na buffer, pH 6.5. The cells, suspended at a concentration of $10^8$ cells per ml in a ice-cold 20mM Mes-Na buffer, pH 6.5 supplemented with protease inhibitors (EDTA 2mM , Pefabloc 1mM, N-ethylmaleimide (NEM) 1mM, leupeptine and pepstatin 5mg/ml each), were broken in a cell cracker [18]. The post nuclear supernatant was centrifuged (100000g, 1 h, 4°C) in order to separate cytosolic and membrane proteins. Separation of integral membrane proteins was then essentially conducted as described by Bordier [19]. The pellet (membrane fraction) was solubilized in 10 ml of Tris-HCl 10 mM , pH 7.4, NaCl 150 mM, and Triton X-114 1% at 0°C. The sample was centrifuged (58000g, 20 min, 0°C) to pellet the insoluble fraction. In a conical tube, the supernatant was overlaid on a sucrose cushion (6%, w/w), containing Tris-HCl 10 mM , pH 7.4, NaCl 150 mM and Triton X-114 0.06% (30 ml) and incubated for 3 min at 30°C. The tube was centrifuged (300 g, 3 min, 30°C) using a swinging bucket rotor. The integral membrane protein rich phase was recovered in the detergent rich phase at the bottom of the tube. The amount of integral membrane protein present in the sample (about 2 mg) was determined by BCA method [16].

Tubulin

Microtubule proteins were prepared as previously described [20]. Pure tubulin was isolated from microtubule proteins by phosphocellulose chromatography in MME buffer (Mes 0.1M pH 6.75, $MgCl_2$ 1mM, EGTA 1mM) as previously described [21]. Phosphocellulose-purified tubulin was diluted directly in the required denaturing solution and used in 2-D electrophoresis experiments.

Isoelectric focusing: immobilized pH gradients

Gel casting

The Immobiline concentrations used to generate the pH gradients (mean buffering power = 3 mequiv.l-1.pH-1) were calculated according to published recipes [22]. Linear pH gradients with plateaus were used. The acrylamide percentage varied from 3.2 to 3.5%T for sample application by rehydration [10]. For casting the low percentage gels, the previously described modifications [10] were used.

IPG rehydration solutions

Various denaturing cocktails were tested. The standard cocktail contained 8M urea, 4% CHAPS, 0.5% Triton X100, 0.4% carrier ampholytes (Pharmalytes 3-10) and 10mM DTT.

For evaluation of the efficiency of the denaturing cocktails, the following parameters were investigated, either alone or in combination.

-alteration of the ampholytes concentration (from 0.4 to 1.6%).

-alteration of the detergent mix used. Various zwitterionic detergents were tested, alone or in combination with CHAPS (2%) and Triton X100 (0.5%). These detergents included amidosulfobetaines C11 and C12 at a concentration of 2% [23], classical linear sulfobetaines (SB3-10 and SB3-12) at a concentration of 2%. As the latter detergents are not compatible with high concentration of urea, the urea level was reduced to 5M (with SB 3-10) or 4M (with SB3-12). In some experiments, the non-detergent sulfobetaine NDSB 256 (benzyl dimethyl ammonio propane sulfonate) was added (5% final concentration) to a CHAPS-based denaturing cocktail.

-alteration of the chaotrope used. Various mono-substituted alkyl ureas (from methyl to propyl urea) were used either alone (7 to 8M final concentration) or in combination with urea. Butylurea and thiourea could only be used in combination with urea. Concentration up to 1M (butylurea) or up to 3M (thiourea) were tested.

Dissolution of the chaotropes was achieved by gentle heating (37°C) and vortexing. Most of the denaturing solutions are clear at 20°C, but precipitate in the cold. The solutions can be kept frozen at -20°C, and are gently warmed (37°C) to redissolve the chaotropes before use.

The most representative denaturing cocktails are presented in figure legends and in the results and discussion sections.

IPG strip rehydration and running

4mm wide IPG strips were cut from the dry plate with the use of a paper cutter. They were rehydrated in grooved chambers, as previously described [10]. The protein sample was added to the denaturing solution to a final volume of 400 to 500 µl. Rehydration was allowed to proceed overnight to ensure maximal diffusion of the proteins within the strip.

The first dimension strips were run on a Dry-strip kit according to the manufacturer's instructions with the previously described modifications [10]. The entire set-up was covered with low viscosity silicon or paraffin oil, and the temperature set at 22°C with a circulating water bath.

Migration was carried out at 150V for 30 minutes, 300V for 2H30, then ramping to 3400V within one hour and finally 3400 V to reach a total of 50 to 60 kVh. The necessity of the low voltage steps was not investigated.

Equilibration

After the IEF run, the oil was poured off, and the strips equilibrated while in place in the running setup [24]. The strips were then sealed on the top of the 1.5mm thick second dimension gel (Bio-Rad vertical system) with the help of 1% low-melting agarose in 0.2% SDS, 0.15M BisTris-0.1M HCl buffer supplemented with bromophenol blue as a tracking dye. In some experiments, the composition of the equilibration solution was altered. The urea concentration was varied between 6 and 9M, the SDS concentration was varied between 2.5 and 5%. An equilibration solution containing 1M butylurea or 2M thiourea in addition to urea was also tested.

Isoelectric focusing: carrier ampholyte-based gradients

Gel casting, running and equilibration

Tube gels (3 mm diameter, 160 mm long) were cast in cut 1 ml disposable glass pipets (Corning). The gel (4.5%T, 2.7% C-PDA) also contained 9M urea, 4% CHAPS, 0.4% Triton X100, 2% Resolyte 4-8 and 0.4% Pharmalyte 3-10. Polymerization was initiated by addition of 1µl of TEMED and 4µl of 10% persulfate per ml of gel mixture, and allowed to proceed for 2 hours at room temperature. The sample was applied at the cathodic end. The anodic electrolyte was 15mM phosphoric acid and the cathodic electrolyte was 20mM degassed sodium hydroxide. The gels were run in a Chamber model 175 (Bio-Rad) at 250V for 2 hours, 500 V for 2 hours and 1500V for 15 hours (total 24 kVh).

After the run, the gels were removed from the running tube by water pressure. The first cm of the gel (anode) was dipped for ≈ 10 seconds into a solution of pyronin Y (1%) before complete expulsion of the gel into the tube containing the equilibration buffer (2.5% SDS, 0.15M BisTris-0.1M HCl, 2mM tributylphosphine, 20% glycerol). The gels were equilibrated for 20 minutes, loaded onto the SDS gels and sealed in place with agarose in stacking buffer (0.15M BisTris-0.1M HCl) containing 0.2% SDS.

SDS PAGE

The gels used in this study were continuous 10% T 2.7%C gels using PDA as a crosslinker and thiosulfate included in the gel formulation to reduce background upon silver staining [25]. The gel buffer (final concentration) was 0.167M Tris-0.1M HCl. A 5mm-wide stacking gel (4%T) was cast in the 0.15M BisTris-0.1M HCl buffer. The upper electrode buffer (made fresh every time) was 0.1M Tris, 0.1M taurine, 0.1% SDS, while the lower electrode buffer was the standard Tris-glycine-SDS buffer (twice concentrated) and was kept in the electrophoresis cell for ca. 3 months at 4°C. Migration was carried out at 25V for 1 hour and then at 35-40mA/ gel for 5-6 hours or 15mA/gel overnight, until the tracking dye reached the bottom of the gel.

Staining

Depending upon the purpose of the gel (analytical or micropreparative) different staining protocols were used.

-For analytical gels (25-200 μg of a complex sample loaded in the first dimension) the tetrathionate-silver nitrate staining was used [10].

-For low-sensitivity silver staining (250-500 μg of proteins loaded on the gels) the gels were stained with ammoniacal silver as described by Hochstrasser [25] except that the gels were treated neither with glutaraldehyde nor with any other sensitizer in order to avoid saturation of the silver staining by excessive sensitivity.

-For micropreparative scale (>500 μg of proteins loaded on the gels), the gels were stained for 24 hours by colloidal Coomassie blue G [26] with the modifications described by Anderson [27].

RESULTS

Preliminary experiments

2-D PAGE of membrane proteins with IEF in immobilized pH gradient in the first dimension was shown to be plagued with quantitative losses of proteins, while the resolution is kept high [14]. This phenomenon has been attributed to insolubilization of proteins at, or very close to, their isoelectric point in the IPG matrix. We first tried to alleviate this problem by changing the composition of the equilibration solution. We increased the SDS concentration (up to 5%), and the urea concentration (up to 9M).We also added butylurea (1M), as this compound has been claimed to improve the solubility of membrane proteins [28]. Typical results are shown in Figure 1, and indicate that none of these approaches allowed to increase significantly the amount of membrane proteins present in the 2-D gel. The addition of higher amounts of carrier ampholytes in the focusing gel (up to 1.6%) did not bring any improvement, as stated earlier [14].

In a second approach, we tried to improve the solubility of the proteins by altering the detergents used in the first dimension. To this purpose, we tested polyol-based detergents (octylglucoside [29], and methylglucamine-based detergents [30]), we reinvestigated the long chain amidosulfobetaines [23], and we also tested a compound with two dipolar heads (myristylamido-bis (propyl dimethyl ammonio propane sulfonate), synthetized as previously described [31]. These compounds were tested either alone or in combination with CHAPS and Triton X100. They were not able to increase the amount of protein present in the 2-D gel (data not shown).

In a third step, we tried to use conventional linear sulfobetaines. These detergents are not compatible with high amounts of urea [23], so that we tried to replace at least part of the urea by another chaotrope. We first tested chaotropes which were claimed to be at least as efficient as urea, namely substituted ureas [32], formamide and thiourea [33]. Typical results are shown in Figure 2. It can be easily shown that formamide and methylurea are not efficient and induce a severe loss in resolution (Fig 2B-C). Similar results were obtained with other alkylureas up to propylurea, and with dimethyl urea (data not shown). The intensity of streaking is parallel to the concentration of the alkylurea and therefore slightly less severe when an urea-alkylurea mixture is used. However, none of these alkylureas was associated to a high resolution, with any of the detergents tested.

On the other hand, thiourea, used in combination with urea since its solubility in water is too low, was compatible with a high resolution (Fig 2-D). The urea-thiourea cocktail seemed to be as least as efficient as the control (Fig 2-D vs 2A). Our subsequent experiments were based on the evaluation of the usefulness of thiourea for the 2-D electrophoresis of sparingly-soluble proteins.

Tests on nuclear proteins

Nuclear proteins were the first type of proteins tested. The purpose of this test was to find the best combination between urea, thiourea and detergents and to test whether any of these combinations was superior to the standard conditions (urea 8M, CHAPS 4%). As many nuclear proteins are moderately soluble, high loads of proteins are required to see an improvement in the solubility of these proteins. We therefore used medium-range loadings of proteins (≈ 500µg), coupled with a low-sensitivity silver staining. The results are shown on figure 3. It can be seen that detergents weakly compatible with urea, as SB 3-12, do not yield an acceptable resolution (Fig 3B-C). On the contrary, a cocktail containing 2M thiourea, 5M urea and 2% SB 3-10 gave a gel of adequate resolution (Fig 3E), and an improvement in the solubility of some proteins could be noted. To test whether this improvement was mainly due to the detergent or to the addition of thiourea, we tested in parallel a cocktail containing 5M urea and 2% SB 3-10, but no thiourea and a cocktail containing 7M urea, 2M thiourea and 4% CHAPS. The resulting gels are shown in Fig 3D and 3F, respectively. Here again, the resolution was adequate and the solubility was improved in the presence of thiourea compared to the control (Fig 3F vs 3D and 3A). From these experiments, it appeared that the presence of 2M thiourea in the IPG gel seemed to be the key factor in the increase of solubility. We therefore tried to vary the thiourea concentration from 1 to 3M, by increments of 0.5M. 2M thiourea was the optimum, lower concentrations giving a lesser solubility, while concentrations of 2.5M and above led to some streaking (data not shown).

Tests on membrane proteins

We continued our tests on membrane-associated proteins, mainly microsomal proteins. We reasoned that membrane proteins could be more sensitive to the nature of the detergent than nuclear proteins, so that we kept testing in parallel two thiourea-containing solutions: one with 5M urea and 2% SB 3-10, and one with 7M urea and 4% CHAPS.

As thiourea was found to be an important enhancer of the solubility of proteins, we tested whether its presence would be beneficial in the equilibration solution in order to keep the proteins soluble while they are saturated with SDS. The results, displayed in Figure 4, show on the contrary that the presence of thiourea in the equilibration solution impairs the quality of the resulting 2-D gel (vertical streaking), whether the IPG gel contains thiourea or not. Standard equilibration was therefore kept for subsequent experiments.

We then tested the improvement in solubility brought by thiourea on a selected population of sparingly-soluble proteins, namely integral membrane proteins. These proteins can be readily isolated by their selective partition in the detergent-rich phase of a Triton X114-based two-phase system [19]. Integral membrane proteins were analysed by 2-D electrophoresis with IPG in the first dimension with various denaturation cocktails, containing thiourea or not. The results are displayed on Figure 5 and show that thiourea brings a major improvement in the solubility of these proteins. However, the total concentration of chaotropes seems to be the key parameter, overwhelming the nature of the detergent. This can be seen from the fact that integral membrane proteins, especially of high molecular weight, are more efficiently solubilized by a solution containing CHAPS (a mild detergent), 2M thiourea and 7M urea than by a solution containing SB 3-10 (a strong detergent), 2M thiourea but only 5M urea (Fig. 5 C vs 5B).

In order to test whether additional compounds could further increase the solubility of proteins in thiourea-containing denaturing solutions, we used a variety of amphiphilic sulfobetaines, added to a thiourea-urea-CHAPS cocktail. The results on microsomal proteins are shown on Figure 6. Figure 6A shows the 2-D electrophoretic pattern of microsomal proteins obtained with carrier ampholytes-supported IEF. The resolution is rather poor, but the solubility of proteins is good, as shown by the presence of massive spots of high (> 50kDa) molecular masses. On the contrary, the resolution is much higher when IPG was used for focusing, but the quantity of these high molecular mass spots decreased severely (Figure 6B). When different thiourea-containing solutions were tested (Fig 6C-F), the solubility of the proteins increased, but high molecular mass spots were still strongly under-represented. However, the addition of 0.2M benzyldimethylammonio propane sulfonate [34] (NDSB 256 ) notably improved the solubility of high molecular mass spots (Fig 6F). This is attributed to the fact that this amphiphilic sulfobetaine exerts a detergent-like action which reinforces the action of CHAPS and thus increases the solubility of proteins. However, the solubility of some proteins is still lower in this optimized IPG system than in conventional carrier ampholytes-supported IEF.

Tubulin

Tubulin is a cytoplasmic protein which is highly polymorphic and highly prone to aggregation. Because of its numerous functions in the cell, this protein has been extensively studied. Tubulin undergoes a variety of post-translational modifications, including the addition of amino acids [35]. For these reasons, the analysis of tubulin by 2-D electrophoresis with shallow pH gradients and at high loads for subsequent detailed analysis is a very important tool in the field [36]. However, when we tried to use immobilized pH gradients for the analysis of tubulin at high loads, disappointing results were obtained, with very poor penetration of the protein (Fig 7A). This was attributed to the aggregating properties of tubulin, leading to almost irreversible insolubilization at the isoelectric point. Here again, the use of thiourea dramatically improved the penetration of the protein in the 2-D gel (Fig 7B-C).

DISCUSSION

Two-dimensional electrophoresis with immobilized pH gradients in the first dimension has now proven its efficiency in terms of resolution and reproducibility [6, 37, 38]. Moreover, this technique allows a very easy scale-up, so that micropreparative gels are easily carried out with minimal loss in resolution [8-10], which is not the case for 2-D gels using carrier ampholytes-based focusing. While these excellent features are true for soluble proteins (e.g. biological fluids) or for whole cell extracts where soluble proteins dominate, some problems are experienced when less soluble proteins are analysed [14]. In this case, severe quantitative losses are observed, which precludes the use of IPG when micropreparative work is planned. These losses were attributed to adsorption to the matrix at the isoelectric point [14].

In order to alleviate this problem, we first attempted to increase the solubilizing power of the equilibration solution, but this approach proved unsuccessful. We therefore tried to increase the solubility of the proteins in the focusing dimension. To this purpose, we varied the detergents and the chaotropes used for rehydrating the IPG strips. We tried to select the most powerful detergents and chaotropes available for IEF, i.e. without any net charge.

As we suspected that hydrophobic interactions were the major cause of protein losses, we first tried to use powerful detergents. The most powerful ones are linear zwitterionic detergents [39]. However, these detergents form inclusion compounds with urea [23], so that high concentrations of urea cannot be used. This led us to try to replace part or the whole of urea by other chaotropes. Here again, we tried to select chaotropes as powerful as possible. This led us to test mainly alkylureas [32] and thiourea. The latter compound is weakly soluble in water (ca. 1M), but is more soluble in urea solutions. Such urea-thiourea solutions have been shown to be very efficient denaturants [33], [40]. Our results clearly show that alkylureas dramatically hamper resolution, while urea-thiourea mixtures gave excellent results and improved the solubility of proteins in many cases (see Results). It must be emphasized that the extent of effects shown in the figures is far beyond the effects due to gel to gel variability. Optimal transfer in the second dimension was however obtained with an equilibration solution devoid of thiourea. Chaotropes are known to weaken the SDS-protein interactions [41], and it is likely that thiourea, when present in the equilibration solution, hampers proper saturation of the proteins with SDS. When a thiourea-free equilibration solution is used, the thiourea present in the strip increases the solubility of the proteins during the IEF step but diffuses out during the equilibration process, so that proper saturation of proteins with SDS diffusing into the strip is achieved. This double diffusion process probably explains why thiourea concentrations higher than 2M are less efficient.

The ideal conditions would of course combine the highest chaotropic power, i.e. 2M thiourea and 7 to 8M urea, with a detergent cocktail as efficient as possible. In the current state of commercially available detergents, this is however not possible. Detergents compatible with high concentrations of urea (e.g. CHAPS or SB 3-8) are weakly efficient [39], while detergents with longer tails are more efficient but not compatible with high concentration of urea. The best compromise will therefore vary from protein to protein, but two conditions seem worth to be retained. Some proteins, which require an efficient detergent, will be better solubilized in a mixture containing 2M thiourea, 5M urea, 2% CHAPS and 2% SB 3-10. Other proteins, which require a high concentration in chaotropes (e.g. tubulin), will be better solubilized in a mixture containing 7M urea, 2M thiourea, 4% CHAPS and, in some cases, a second amphiphilic sulfobetaine such as NDSB 256 [34].

It must be emphasized that we did not encounter the case of proteins which were better solubilized in the absence of thiourea than in one of the two cocktails mentioned above, so that these solutions can be recommended for general use, and are now routinely used in our laboratory. Owing to the improvement brought by amphiphilic sulfobetaines, our future research effort will be directed to the synthesis and testing of new zwitterionic detergents, in order to be able to conciliate high detergency and maximum chaotrope concentration.

The last point to be emphasized is that thiourea, as many other sulfur containing compounds, strongly inhibits acrylamide polymerization and is therefore not compatible with the standard setup for CA-IEF in tube gels.

Acknowledgments:
The gift of chloroplast envelope by Christine Miege and Catherine Albrieux is gratefully acknowledged. Grateful thanks are also due to Bernard Eddé for stimulating discussions about tubulin separation by 2-D electrophoresis.

Figures

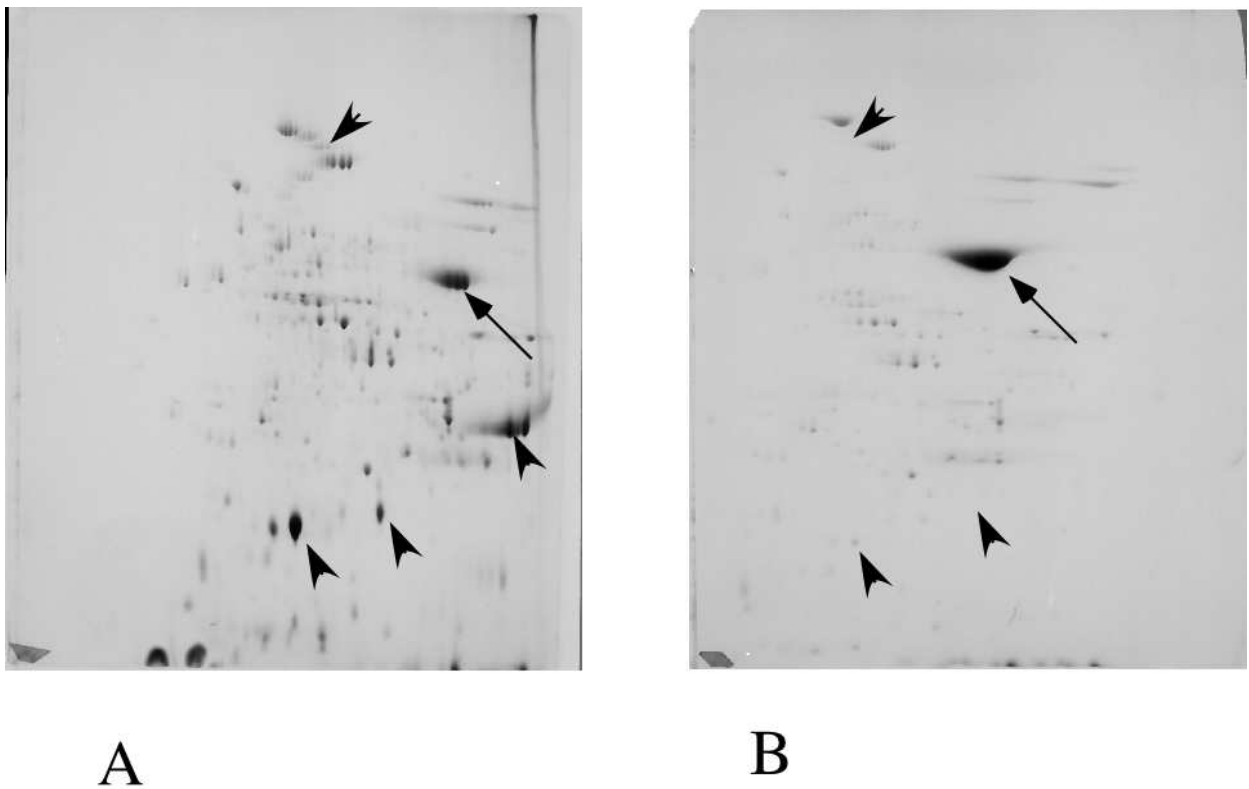

A				B

Figure 1. 2-D electrophoresis of chloroplast enveloppe proteins (200µg). Detection with silver staining.
A: IEF with carrier ampholytes (pH gradient 4 to 8). B: IEF with IPG (linear 4-8 pH gradient).
The arrow points to RUBISCO (large subunit). Although the pH gradient in the case of CA-IEF is severely drifted, many proteins are present at higher final quantities in the gel with CA-IEF than in the gel with IPG. Some of these IEF mode-sensitive proteins are indicated with arrowheads.

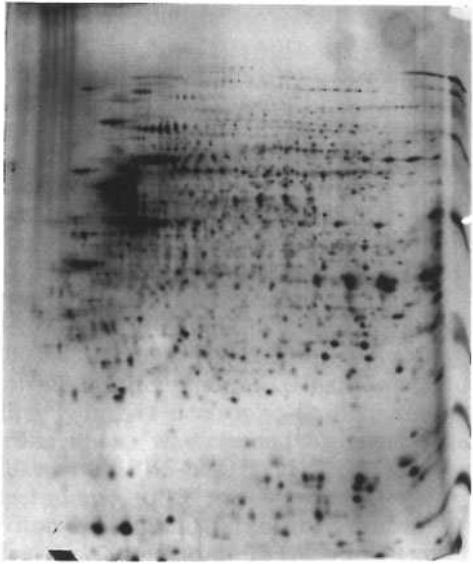 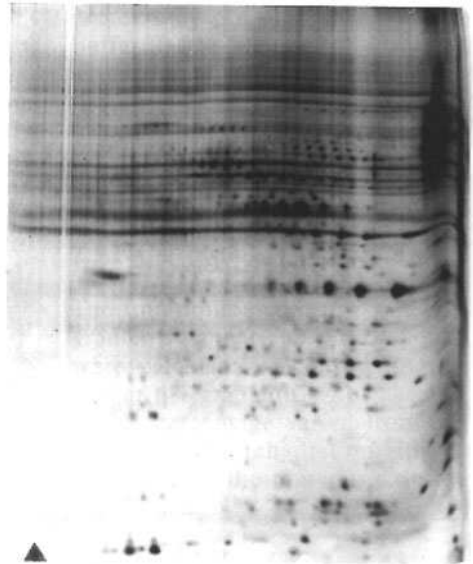

A  B

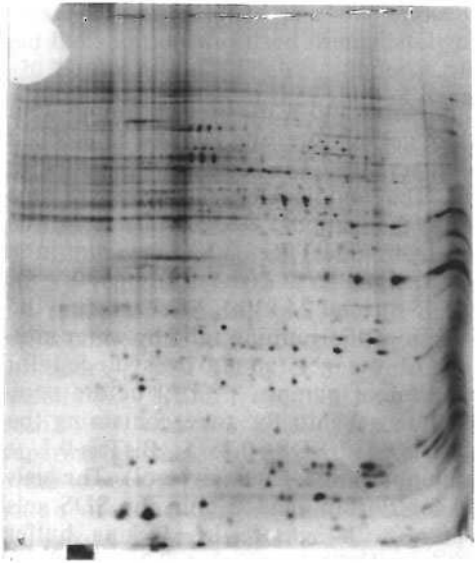 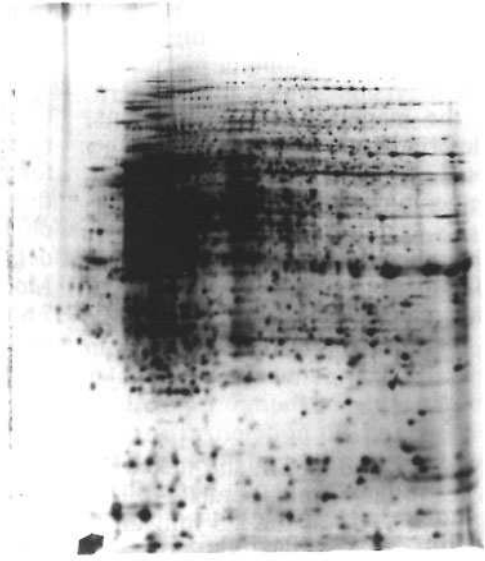

C  D

Figure 2. 2-D electrophoresis of total cell extract proteins from P3-X63-Ag8 cells (100μg). IEF with IPG (linear 4-8 pH gradient). Detection with silver staining. In addition to the chaotrope and detergent, the IPG solution also contained the following secondary additives: 0.5% Triton X100, 0.4% carrier ampholytes (Pharmalytes 3-10) and 10mM DTT.
A: IPG in 8M urea, 4% CHAPS; B: IPG in 50% (v/v) formamide, 2% SB 3-10; C: IPG in 8M methylurea, 2% SB 3-10; D: IPG in 2M thiourea, 5M urea, 2% SB 3-10.

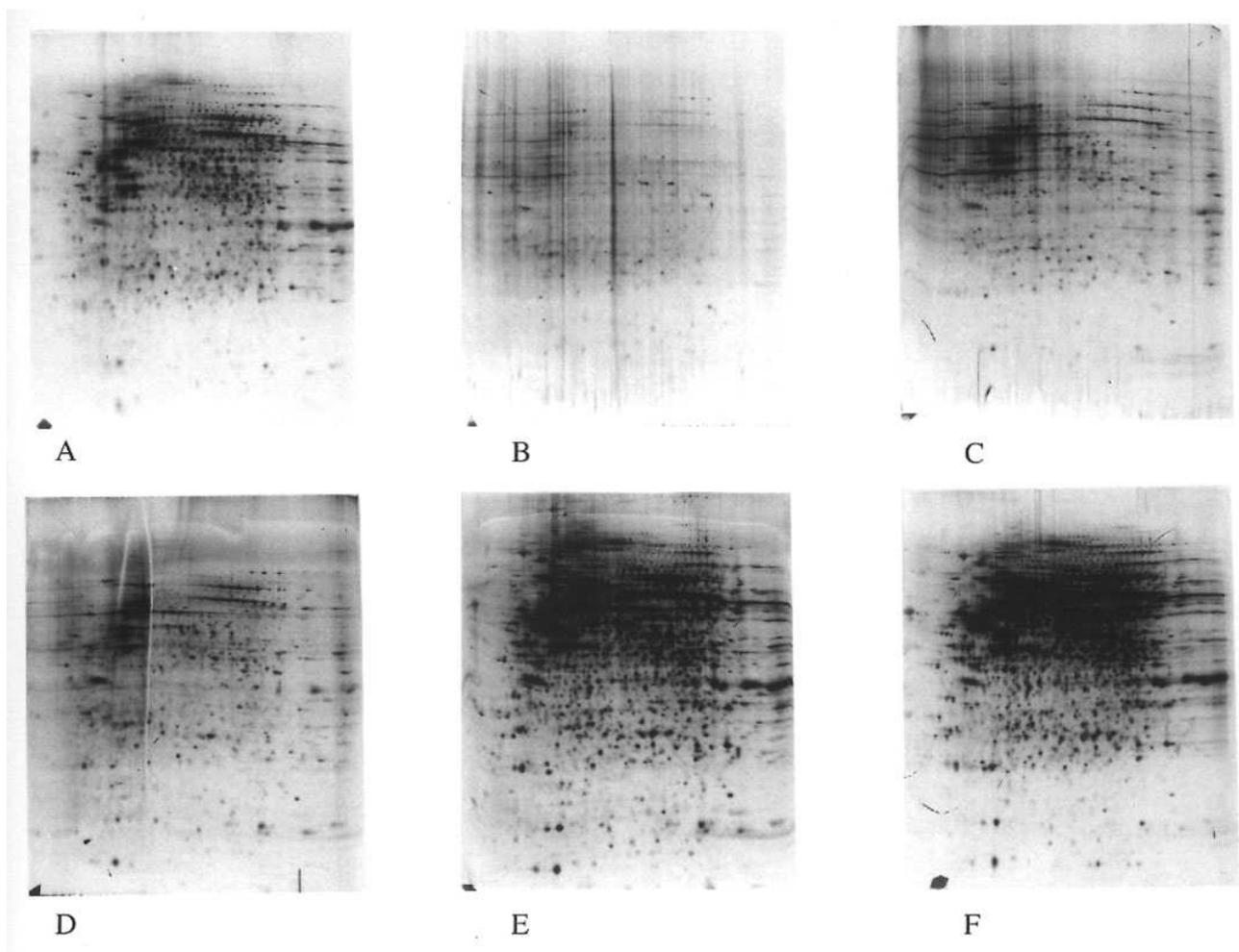

Figure 3. 2-D electrophoresis of nuclear proteins from C2C12 cells (500μg). IEF with IPG (linear 4-8 pH gradient). Detection with low-sensitivity silver staining. Secondary additives as in Figure 2.
A: IPG in 8M urea, 4% CHAPS; B: IPG in 4M urea, 2% SB 3-12; C: IPG in 4M urea, 2M thiourea, 2% SB 3-12; D: IPG in 5M urea, 2% SB 3-10; E: IPG in 2M thiourea, 5M urea, 2% SB 3-10; F: IPG in 2M thiourea, 7M urea, 4% CHAPS.

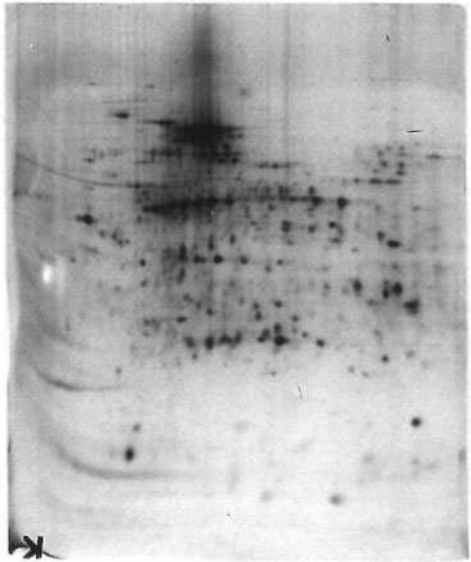 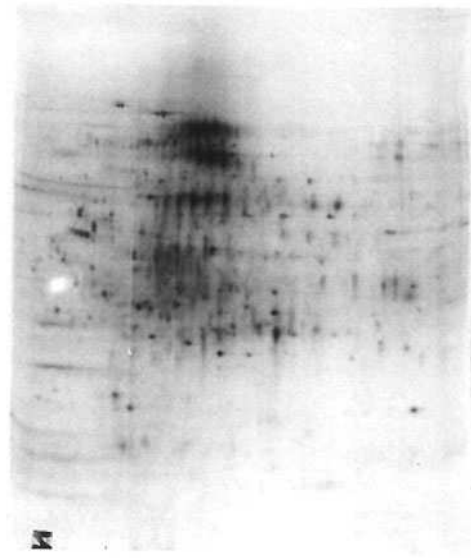

A B

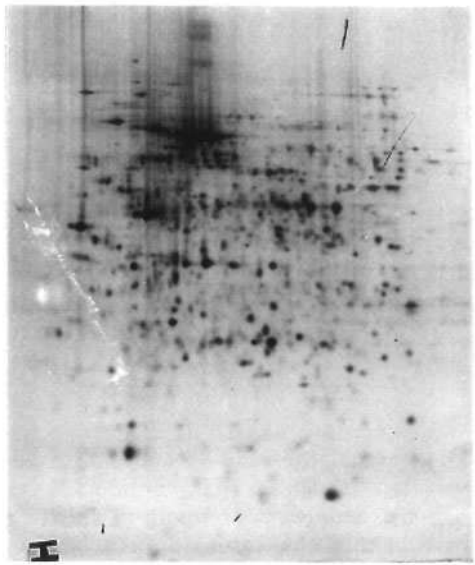 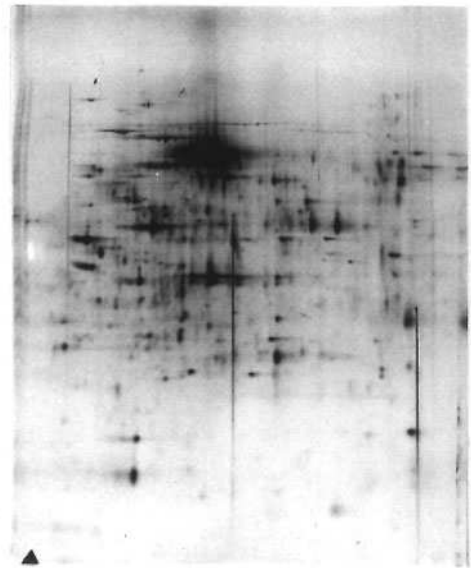

C D

Figure 4: 2-D electrophoresis of microsomal proteins from P3-X63-Ag8 cells (100µg). IEF with IPG (linear 4-8 pH gradient). Detection with silver staining. Secondary additives as in Figure 2.
A: IPG in 8M urea, 4% CHAPS, equilibration in standard solution; B: IPG in 8M urea, 4% CHAPS, equilibration in standard solution + 2M thiourea; C: IPG in 5M urea, 2M thiourea, 2% SB 3-10, equilibration in standard solution; D: IPG in 5M urea, 2M thiourea, 2% SB 3-10, equilibration in standard solution + 2M thiourea.

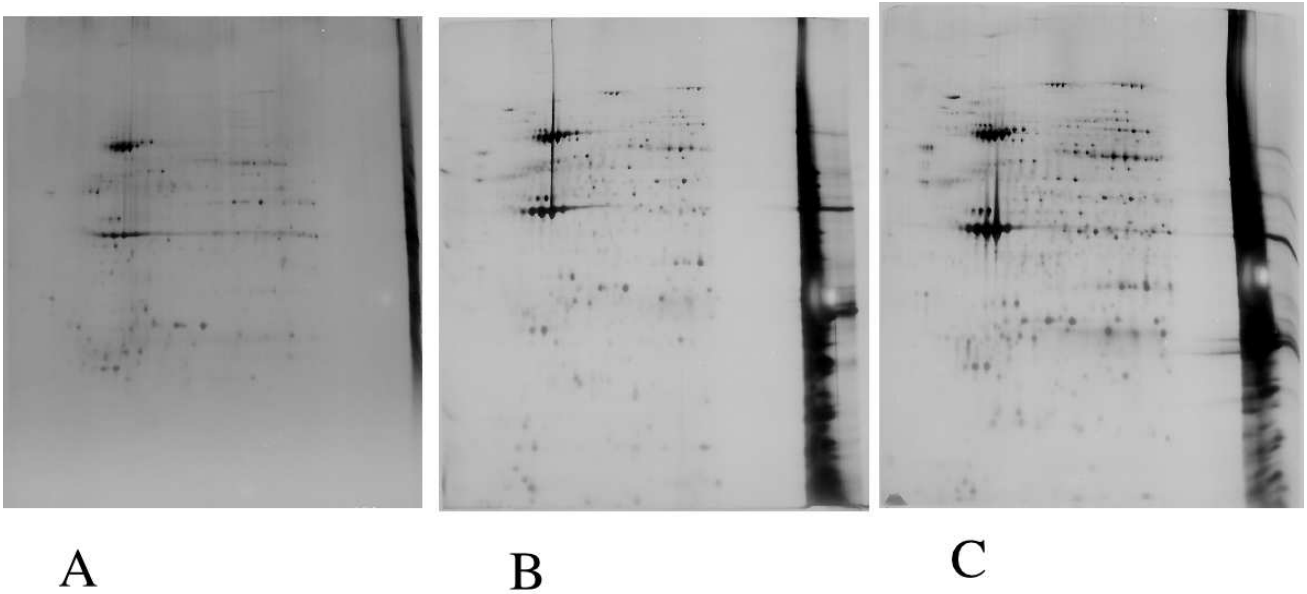

Figure 5: 2-D electrophoresis of integral membrane proteins from *D. discoideum* (100μg). IEF with IPG (linear 4-8 pH gradient). Detection with silver staining. Secondary additives as in Figure 2.
A: IPG in 8M urea, 4% CHAPS; B: IPG in 2M thiourea, 5M urea, 2% SB 3-10; C: IPG in 2M thiourea, 7M urea, 4% CHAPS.

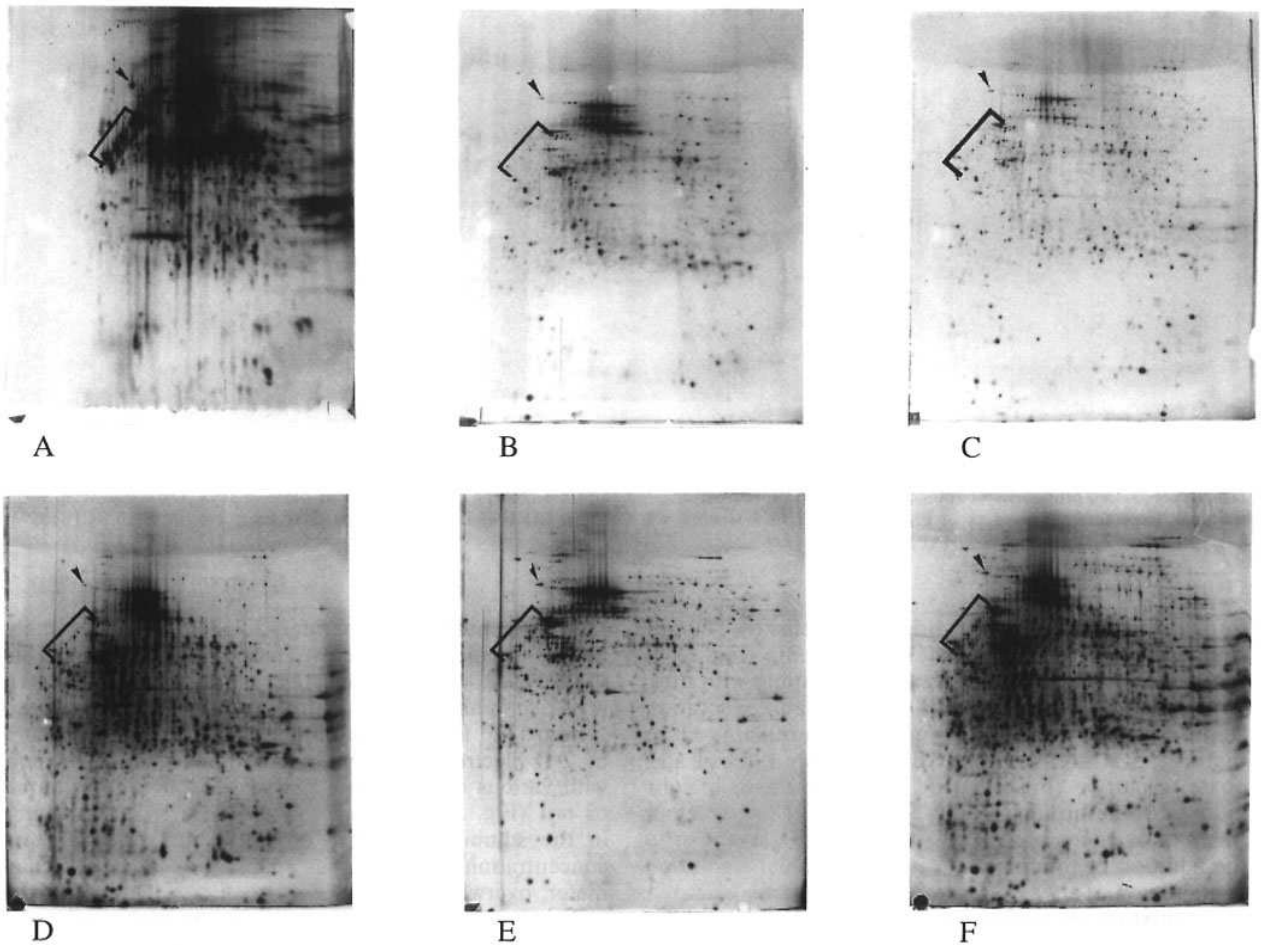

Figure 6: 2-D electrophoresis of microsomal proteins from C2C12 cells (200μg). IEF with IPG (linear 4-8 pH gradient). Detection with silver staining. Secondary additives as in Figure 2.

A: IEF with carrier ampholytes; B: IPG in 8M urea, 4% CHAPS; C: IPG in 5M urea, 2% SB 3-10; D: IPG in 5M urea, 2% SB 3-10, 2% CHAPS; E: IPG in 2M thiourea, 7M urea, 4% CHAPS; F: IPG in 2M thiourea, 7M urea, 4% CHAPS; 5% NDSB 256. The bracket and arrowhead indicate high molecular weight proteins prominent when CA-IEF is used, and under-represented in IPG.

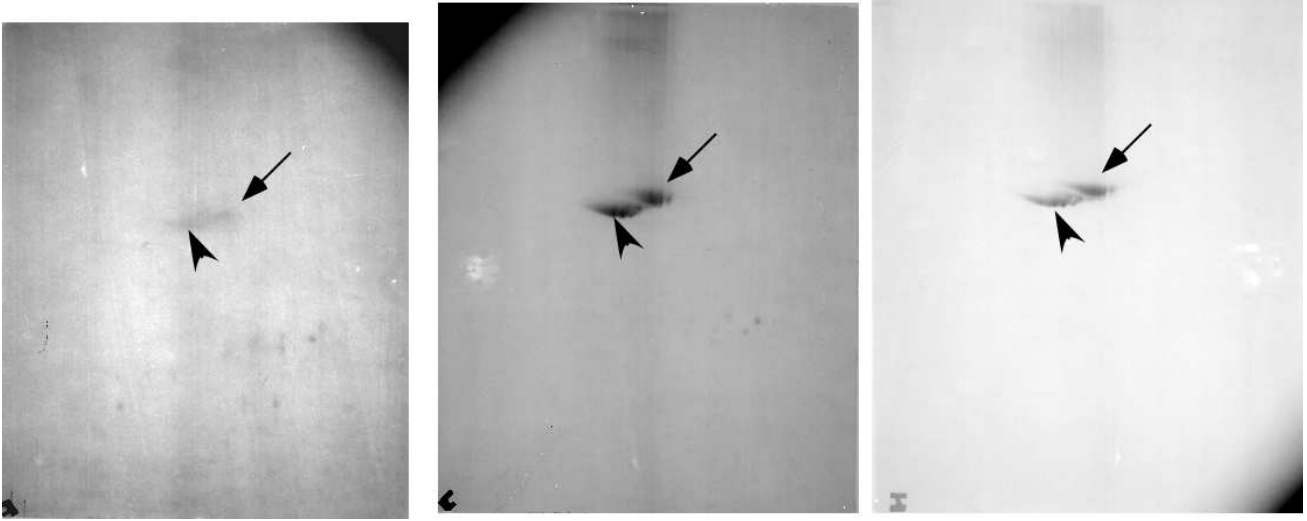

Figure7: 2-D electrophoresis of purified bovine brain tubulin. IEF with IPG (linear 4-6 pH gradient). Detection with Coomassie Blue staining. Secondary additives as in Figure 2
A: IPG in 8M urea, 4% CHAPS; B: IPG in 2M thiourea, 5M urea, 2% SB 3-10; C: IPG in 2M thiourea, 7M urea, 4% CHAPS.
arrow: α tubulin, arrowhead: β tubulin